\documentclass[showpacs, twocolumn]{revtex4-1}
\usepackage{graphicx}
\usepackage{amsmath}
\usepackage{appendix}

\begin{document}

\title{ Effects of the Lee-Huang-Yang quantum corrections on a disordered dipolar Bose gas}

\author{Abdel\^{a}ali Boudjem\^{a}a$^{1,2}$}
\affiliation{$^1$Department of Physics, Faculty of Exact Sciences and Informatics, 
and $^2$Laboratory of Mechanics and Energy, Hassiba Benbouali University of Chlef, P.O.
Box 78, 02000, Ouled-Fares, Chlef, Algeria.}
\email {a.boudjemaa@univ-chlef.dz}


\begin{abstract}

We study the behavior of a quantum dipolar Bose condensate with the Lee-Huang-Yang  quantum corrections at zero temperature in the presence of weak disorder potential. 
We solve the underlying nonlocal Gross-Pitaevskii equation using a perturbative theory.
The role of both the Lee-Huang-Yang term and the disorder potential on the condensed fraction, the equation of state, the compressibility 
and the superfluid density is deeply analyzed.
Surprisingly, we find that the Lee-Huang-Yang  quantum corrections not only arrest the dipolar implosion but also lead 
to reduce the condensate fluctuations  due to the disorder effects prohibiting the formation of a Bose glass state.
We show that the superfluid density exhibits unconventional behavior.

\end{abstract}

\maketitle

\section{Introduction }

The interacting Bose gas in a weak random external potential represents an interesting model for studying the relation between 
Bose-Einstein condensation (BEC) and superfluidity and has been the subject of many experimental and theoretical investigations in the past two decades. 
Experimentally, the so-called dirty boson problem was first studied with superfluid helium in aerosol glasses (Vycor) \cite{Crook, Chan, Rep}.  
Recently, disordered BECs trapped in optical potentials have been investigated extensively by many groups \cite{Clm,Schut, Lye, Clm1, Bily, Roat, Chen, Wht}. 

On the theory side, the dirty boson problem was first considered by Huang and Meng employing theùù Bogoliubov theory \cite{HM}
in order to understand the transition mechanism from superfluid to Bose glass phase. 
Later on this approach was generalized utilizing the original framework of second quantization \cite{Gior, Mish,Lugan2, Falco,Yuk} and the replica method \cite{Lopa, Zob, Mor, Bhong}.
Alternatively, in Refs \cite{LSP, Lugan, Lugan1, Gaul, Gaul1,Lell} the disorder averaging is implemented by the perturbative theory.
The main finding emerging from the above studies is that the condensate fragments into low-energy localized particles 
due to the localization of bosons in the respective minima of the external random potential, thus, lowering the superfluid flow.

In the context of ultracold atoms with dipole-dipole interactions (DDI), the properties of 
disordered uniform BECs have  aroused a high interest in many area of research \cite{Krum,Nik, Ghab, Boudj,Boudj1,Boudj2, Boudj3}. 
In three-dimensional (3D) geometry, it has been found that the superfluid density acquires a characteristic direction dependence due to the anisotropy of the DDI  
\cite{Krum,Nik, Ghab, Boudj,Boudj1}.
In 2D case,  we have shown that the interplay between disorder and rotonization induced by the DDI may strongly reduce
the superfluidity leading to the transition to a superglass state \cite{Boudj2, Boudj3}.

Quantum fluctuations which was first introduced by Lee-Huang-Yang (LHY) \cite {LHY}, originate from the zero-point motion 
of the Bogoliubov excitations. They are universal since they depend only on the two-body scattering length. 
The LHY quantum corrections may shift the ground state of the condensate,
as recently observed in strongly contact-interacting Fermi \cite {Alt,YI,Nav} and Bose gases \cite {Pap,Nav1}.
In the case of a dipolar Bose gas, the first-order LHY corrections term scales as  
$(32/3) g n\sqrt{na^3/\pi} {\cal Q}_5(\epsilon_{dd})$ \cite{lime, Boudj4,Boudj5},
where $n$ is the gas density and the functions ${\cal Q}_j(x)=(1-x)^{j/2} {}_2\!F_1\left(-\frac{j}{2},\frac{1}{2};\frac{3}{2};\frac{3x}{x-1}\right)$,
reach their maximal values at $x= 1$ and become imaginary for $x>1$.
It has been revealed that such LHY corrections provide an extra repulsive term arresting the dipolar collapse 
resulting in the formation of quantum droplets, novel state of matter \cite {Pfau1, Pfau2, Pfau3, Chom, Wach, Bess2, Boudj6}. 
A similar scenario holds also in binary BECs, where the self-repulsive LHY term may compensate the interspecies contact attraction
leading to a stable quantum droplets \cite {Petrov, PetAst, Capp, Cab} in dilute Bose-Bose mixtures.

In this paper we study the effects of the LHY quantum corrections on a dipolar Bose gas subjected to a weak random potential 
with a 3D isotropic laser speckle autocorrelation function. 
Our analysis is based on the perturbative theory which has been proved to be an efficient tool towards a quantitative description of
dirty Bose systems with both short- and long-range interactions \cite{LSP, Lugan, Lugan1, Gaul, Gaul1, Lell, Krum,Nik}. 
Understanding the complex interplay of disorder, DDI and LHY quantum fluctuations remains a major challenge. 
This is the aim of the present article.
We calculate the ensemble-averaged disorder by solving perturbatively up to second order in disorder strength the generalized nonlocal Gross-Pitaevskii equation (NLGPE).
The validity condition of  the developed approach is determined.
The density profile of smoothing solutions are tested by comparing the results computed using direct numerical simulation of the NLGPE equation.
Furthermore, we derive useful analytic expressions for relevant physical quantities, such as the condensed fraction, the chemical potential, the compressibility and the superfluid density.
These results are profoundly discussed, with particular emphasis on the case of correlated laser speckle disorder.
In passing, we recover the properties of disordered dipolar Bose gases as described in our recent paper \cite{Boudj} and in the early Huang-Meng results \cite{HM}. 
We show that the LHY corrections tend to palpably reduce the disorder fraction in the condensate and to a strikingly modify the behavior of the superfluidity. 
In particular, we point out that there is a competition between the disorder, which would drive the bosons into a
localized state, and the repulsive LHY corrections, which tends to delocalize the condensate preventing the occurrence of the insulating phase.

The rest of the paper is organized as follows. In Sec.\ref{flism}, we introduce the subtleties of the perturbation approach which is valid only 
for sufficiently weak disorder potential.
In Sec.\ref{LSP},  we apply our theory to the 3D correlated laser speckle model.
We conclude and outline our future work in section \ref{Concl}.


\section{Model} \label{flism}

Consider a dilute 3D dipolar BEC with a weak random disorder potential $U({\bf r})$ in the presence of the LHY quantum fluctuations 
which may reduce the disorder fraction in the condensate and thus, stabilize the system.
We treat the system in the frame of mean-field theory  and we use the generalized time-independent NLGPE \cite{Pfau2, Wach, Bess2, Wach, Chom, Boudj6}
\begin{align}  \label{GPE}
\mu \psi &=\bigg[-\frac{\hbar^2}{2m} \nabla^2+ U({\bf r}) +\int d{\bf r'} V ({\bf r-r'}) |\psi ({\bf r'})|^2   \nonumber \\
& + g_{\text{LHY}} |\psi|^3 \bigg] \psi,
\end{align}
where $\mu$ is the chemical potential of the condensate. The two-body interactions potential reads
\begin{equation}  \label{DPot}
V({\bf r})=g\delta({\bf r})+\frac{C_{dd}}{4\pi}\frac{1-3\cos^2\theta} { r^3},
\end{equation}
with $g=4\pi \hbar^2 a/m$ corresponds to the short-range part of the interaction and is parametrized by the $s$-wave scattering length $a$. 
The DDI coupling constant $C_{dd}$ is characterized by the dipole-dipole distance  $r_*=m C_{dd}/4\pi \hbar^2$ and
determined by the magnetic moment. Here the dipoles are supposed to be oriented along the $z$-direction, 
and $\theta$ is the angle between ${\bf r}$ and the polarization axis.
The Fourier transform of the potential (\ref{DPot}) is $ V(\mathbf k)=g [1+\epsilon_{dd} (3\cos^2\theta-1)]$, where $\epsilon_{dd}=C_{dd}/3g$.
The last term in Eq.(\ref{GPE}) describes the LHY quantum corrections to the chemical potential.
Its strength is given by $g_{\text{LHY}} \simeq (32/3) g\sqrt{a^3/\pi} (1+3\epsilon^2_{dd}/2)$ \cite{Boudj4,Boudj5, Pfau2, Bess2}, 
where only the lowest order expansion of the function ${\cal Q}_5$ has been taken into account at $\epsilon_{dd}\approx 1$. 
This renders $g_{\text{LHY}}$ real since the imaginary part of ${\cal Q}_5$ is very small \cite{Pfau2}.
Note that the LHY term is evaluated in the frame of the local density approximation.
This means that the disorder is implicitly assumed to be changed smoothly in space on a length scale comparable to the healing length 
or the characteristic correlation length of the disorder.  
Then, the density profile of the condensate follows the modulations of a smoothed random potential.


The disorder potential is chosen to be isotropic and should satisfy the conditions
\begin{subequations}\label {CD}
\begin{align} 
\langle U(\mathbf r)\rangle&=0, \\
\langle U(\mathbf r) U(\mathbf r')\rangle&=R (\mathbf r,\mathbf r').
\end{align} 
\end {subequations}
Here $ \langle \bullet \rangle$ denotes the disorder ensemble average and 
$R (\mathbf r,\mathbf r')$ is the disorder correlation function. 

If the disorder is sufficiently weak, it is possible then to solve the NLGPE (\ref{GPE}) perturbatively in powers of $U$ using the expansion \cite{LSP, Lugan, Lugan1, Gaul, Gaul1,Lell, Krum,Nik} 
\begin{equation}  \label{Exps}
\psi ({\bf r})=\psi_0+\psi_1 ({\bf r})+\psi_2 ({\bf r})+\cdots,
\end{equation}
where the index $i$ in the functions $\psi_i({\bf r})$ signals the $i$-th order contribution with respect to the disorder potential. 
They can be determined by inserting the perturbation series (\ref{Exps}) into the NLGPE  (\ref{GPE}) 
and by collecting the terms up to $U^2$. It is convenient to work in momentum representation: $\psi ({\bf k})=\int d {\bf r} e^{-i \bf k. r} \psi({\bf r}) $
and $U ({\bf k})=\int d {\bf r} e^{-i \bf k.r} U({\bf r})$. \\
The zeroth order gives 
\begin{equation} \label{sers0} 
\psi_0=  \sqrt{\frac {\mu - g_{\text{LHY}} \psi_0^3 } { V({\bf k}=0)}}, 
\end{equation} 
which is the homogeneous solution in the absence of a disorder potential.  \\
The first-order equation reads 
\begin{align}  \label{GPE1}
&\left(\frac{\hbar^2 k^2}{2m} -\mu \right) \psi_1 ({\bf k})+U({\bf k}) \psi_0+ 2 \psi_0^2 V({\bf k}) \psi_1({\bf k}) \\
&+ \psi_0^2 V({\bf k}=0) \psi_1 ({\bf k}) + 4 g_{\text{LHY}}  \psi_0^3 \psi_1 ({\bf k})=0 \nonumber,
\end{align}
yielding the solution (after inserting Eq.(\ref{sers0}))
\begin{equation}   \label{sers1} 
\psi_1({\bf k})= - \frac{ \psi_0} {E_k+2 \psi_0^2 \bar V({\bf k})} U({\bf k}), 
\end{equation}
where $E_k=\hbar^2k^2/2m$  and $\bar V ({\bf k})= V({\bf k})+ (3/2)g_{\text{LHY}} \psi_0$ is the effective interaction between atoms in the condensate.
Expression (\ref{sers1}) which can be regarded as the first order imprint of the potential in the condensate amplitude, 
tells us that the dipolar particles are scattered once by the external potential.
For $E_k \ll \bar V ({\bf k})$, or equivalently $k\xi_d \ll 1$, where  $\xi_d=\xi/\sqrt{\bar V/g}$ with $\xi =\hbar/\sqrt{mg \psi_0^2}$ being the healing length,
the kinetic energy is small and hence, the condensate deformation incurs only external potential effects.
In such a situation the NLGPE  (\ref{GPE}) yields for the total density
$n ({\bf r})= \psi_0^2+n^{(1)} ({\bf r})$, in Fourier representation, $n^{(1)} ({\bf k})=-U({\bf k})/\bar V({\bf k})$. 
In the absence of DDI and LHY corrections, the density reduces to the standard Thomas-Fermi-like shape $n({\bf r})=[\mu-U({\bf r})]/g$.
Whereas, for  $k\xi_d \gg 1$, $U({\bf k})$ is a smoothed potential and thus, the system exhibits slow variations.

The validity criterion of the present perturbation approach requires the condition
\begin{equation}  \label{VC}
U \ll \psi_0^2 \bar V.
\end{equation}
It is clear that the condition (\ref{VC}) is a generalization of the well-known condition ($U\ll \psi_0^2 g$) established in Ref \cite{LSP} for
a disordered BEC with contact interactions. In the absence of the LHY term, the condition (\ref{VC}) reduces to $U \ll \psi_0^2 V$.

The second-order equation can be obtained by using the fact that 
$\int d {\bf r} \psi_1 ({\bf r}) U({\bf r}) e^{-i \bf k.r} =\int \psi_1 ({\bf k- k'})  U ({\bf k'})  d{\bf k'}/(2\pi)^3$. 
After having substituting Eq.(\ref{sers0}), one finds
\begin{align}  \label{GPE2}
&\frac{\hbar^2 k^2}{2m} \psi_2 ({\bf k})+ 2 \psi_0^2 \bar V({\bf k}) \psi_2({\bf k}) \\
&+\int \frac{d \bf k'}{(2\pi)^3}  \bigg \{ U({\bf k-k'}) \psi_1 ({\bf k'})  \nonumber \\
&+ \psi_0 [2\bar V({\bf k'})+ \bar V({\bf k})]  \psi_1 ({\bf k'}) \psi_1 ({\bf k- k'}) \bigg \}=0 \nonumber.
\end{align}
The solution of this algebraic equation gives
\begin{align} 
\psi_2 ({\bf k})&= - \int \frac{d \bf k'}{(2\pi)^3}   \frac{1} {E_k+2  \psi_0^2 \bar V({\bf k})} \bigg\{ U({\bf k-k'}) \psi_1 ({\bf k'})  \label{sers2} \\
&+ \psi_0 \left[2\bar V({\bf k'})+ \bar V({\bf k})\right]  \psi_1 ({\bf k'}) \psi_1 ({\bf k- k'}) \bigg\}.  \nonumber
\end{align} 
This elegant formula clearly shows that the collision processes consists of
double scattering at the external potential $U({\bf r})$ and interaction of two single-scattered particles \cite{Gaul}.
Note that the second and higher-order terms do not really modify the validity condition (\ref{VC}) since their contributions are
negligible as we shall see in Fig.\ref{dens}.a.

The condensate fluctuations due to the disorder potential can be evaluated in terms of 
the disorder ensemble averages as $n_R= \langle \psi ({\bf r})^2 \rangle-\langle \psi ({\bf r}) \rangle^2$, in Fourier space
\begin{equation}  \label{depdis}
n_R= \int \frac{d \bf k}{(2\pi)^3} \frac{n R(\bf k)}{\left[E_k+2 \bar V({\bf k}) n \right]^2}.
\end{equation}
Strictly speaking, the fluctuations term (\ref{depdis}) known also as {\it glassy fraction} originates from 
the accumulation of density near the potential minima and density depletion around the maxima.
For $g_{\text{LHY}}=0$, Eq.(\ref{depdis}) simplifies to that prevailed for a disordered dipolar Bose gas without LHY term \cite{Krum, Nik}.


The shift to the equation of state (EoS) of a dirty dipolar BEC due to the disorder potential at a fixed average particle density is given by 
$\delta\mu = -\int (d {\bf k}/2\pi)^3 E_k |\psi_1 ({\bf k})|^2$. 
We see that the perturbative theory predicts a negative correction to the chemical potential which may affect the sound velocity.
However, the obtained EoS is ultraviolet divergent for any uncorrelated disorder. 
The origin of such a divergency comes from the delta-correlated disorder potential.
To circumvent this issue, we renormalize $\delta\mu$ by introducing $\int d {\bf k}/(2\pi)^3 R({\bf k})/E_k$ \cite{Nik, Boudj, Boudj1}.
Therefore, the total chemical potential reads
\begin{align}  \label{EoS}
\mu&= n V({\bf k}=0) +g_{\text{LHY}} n^{3/2} \\ 
&+ 4 n\int \frac{d \bf k} {(2\pi)^3}  \frac{\left[E_k+\bar V({\bf k}) n \right] \bar V({\bf k}) R(\bf k)} 
{E_k \left[E_k+2\bar V({\bf k}) n \right]^2}. \nonumber
\end{align}
This EoS constitutes a natural extension of that obtained for a disordered dipolar BEC \cite {Nik, Boudj} owing to the extra term provided by the LHY quantum fluctuations. 

The inverse compressibility is defined as $\kappa^{-1} = n^2\partial\mu/\partial n$. 
Then, using (\ref{EoS}), we get 
\begin{align}  \label{Comp}
\frac{\partial\mu}{\partial n}&=V({\bf k}=0) + \frac{3}{2} g_{\text{LHY}} n^{1/2}+\\
 &+4\int \frac{d \mathbf k} {(2\pi)^3}  \frac{E_k \left[ \bar V({\bf k})+(3/4)g_{\text{LHY}} n^{1/2} \right] R(\bf k) } 
{\left[E_k+2 \bar V({\bf k}) n\right]^3} \nonumber.
\end{align}
For $g_{\text{LHY}}=0$, the expression (\ref{Comp}) reduces to that found in Refs \cite{Nik, Boudj}.

The Bose condensed fluid in the presence of disorder is splitted into a macroscopic normal component $n_n/n$ and a macroscopic superfluid fraction $n_s/n$.
In the two-fluid model the total momentum ${\bf P (r)}$ of the moving system which is related to the laboratory system by a Galilean transformation ${\bf r'}={\bf r}+{\bf u} t$, 
$t=t'$, is given by ${\bf P}=m {\cal V} (n {\bf v_s}+n_n {\bf v_n})$, where 
${\cal V}$ stands for the volume, $n=n_s+n_n$ is the total density, ${\bf v_s}$ denotes the superfluid velocity, 
${\bf v_n}={\bf u}-{\bf v_s}$ is the normal fluid velocity with ${\bf u}$ being a boost velocity \cite{Nik,Boudj4}. 
We then solve the underlying inhomogeneous NLGPE in the system coordinates using the above perturbative expansion 
(we refer the reader to \cite{Nik,Boudj4, Cord} for more details on the derivation of the superfluid density).
The normal fraction reads:
\begin{equation} \label{sup}
 \frac{  n_n}{n}=\frac{2\hbar^2}{m} \int \frac{d \mathbf k}{(2\pi)^3} \frac{R({\bf k}) \, {\bf k \otimes k}} {E_k[E_k+2n \bar V(\mathbf k)]^2}.  
\end{equation}
This expression is valid for arbitrary disorder correlation function $R({\bf k})$, and effective two-particle interaction $V(\mathbf k)$. 
For systems possessing cylindrical symmetry, say around the $z$-axis, the tensorial superfluid fraction 
separates into a parallel and a perpendicular part defined respectively, as
\begin{equation} \label{sup1}
 \frac{ n_s^{\parallel}} {n}= 1- 4 \int \frac{d k\, d\theta}{(2\pi)^2}\frac{k^4  R(\bf k)}{\left[E_k+2 \bar V({\bf k}) n \right]^2} \sin \theta \cos^2\theta,
\end{equation}
and 
\begin{equation} \label{sup2}
 \frac{ n_s^{\perp}} {n}= 1- 4\int \frac{d k \,d\theta}{ 8\pi^2} \frac{k^4  R(\bf k)}{\left[E_k+2 \bar V({\bf k}) n \right]^2} \sin \theta \sin^2\theta .  
\end{equation} 
When the interactions and the disorder correlation are isotropic i.e. $V(\mathbf k)$ and  $R(\mathbf k)$ are $\theta$ independent, 
the superfluid fraction in both directions reduces to $n_s/n= 1-4n_R/3n$ \cite{HM}.
This result indicates that the normal component of the superfluid is $4/3$ times larger than the condensate fluctuations due to the disorder effects $n_R$.

\section{Laser speckle potential} \label{LSP}

For the sake of concreteness, we will consider optical speckle potential, which is often used with ultracold atoms experiments  \cite{Clm, Clm1, Bily, Roat, Lye,Chen}. 
Experimentally, an isotropic 3D speckle, can be produced as the interference pattern of many wavevectors inside a closed optical cavity \cite{Kuhn}.
Another realization of  3D disordered speckle configuration was proposed in Ref \cite {Abdulaev}, 
where the speckle is formed in the focal point of an empty ellipsoidal optic cavity.

The autocorrelation function of the 3D isotropic laser speckle is given by  $R({\bf r})=R_0|C_A({\bf r})|^2 $ \cite {Abdulaev}, 
where $R_0=U_0^2$ is the disorder strength, and 
\begin{equation} \label{Cordis}
 C_A(y)=\left| \left(3/y^3 \right) \left(\sin y-y \cos y \right) \right|^2,
\end{equation}
where $y=\pi r/\sigma$ with $\sigma$ being for  the correlation length of the disorder.  
The autocorrelation function (\ref {Cordis}) differs from that used in \cite{Kuhn}, $C_A({\bf r})= \text{sinc} (r/\sigma)$, for the 3D isotropic speckle.
In Fourier space, it can be written as \cite {Abdulaev, Boudj}
\begin{equation} \label{CorrS}
|C_A({\bf k})|^2 = \frac{3}{4\pi} (2\sigma)^3[(2\sigma k)^3-12(2\sigma k)+16].
\end{equation}
The function $|C_A({\bf k})|^2$ is normalized by a factor $3 (2\sigma)^3/4\pi$.
An important property of the autocorrelation function (\ref{CorrS}) is that it constitutes the most tractable mathematical model.
Furthermore, $|C_A({\bf k})|^2$ vanishes for $k=1/\sigma$, indicating that the momentum 
only varies in a finite interval from zero, in contrast to the case for a Gaussian function  which 
meets some hindrances in its application to the disordered BEC \cite {Abdulaev, Boudj}.

\begin{figure}
\includegraphics[scale=0.8]{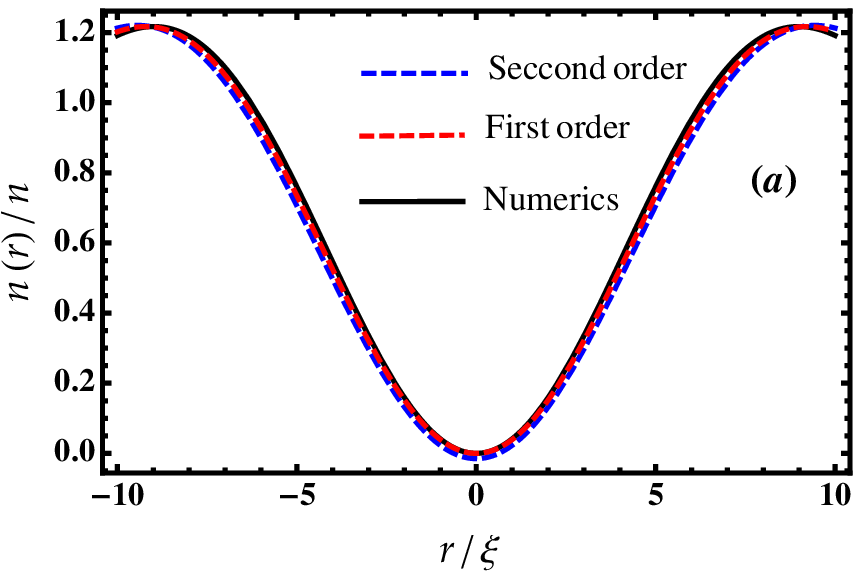}
\includegraphics[scale=0.8]{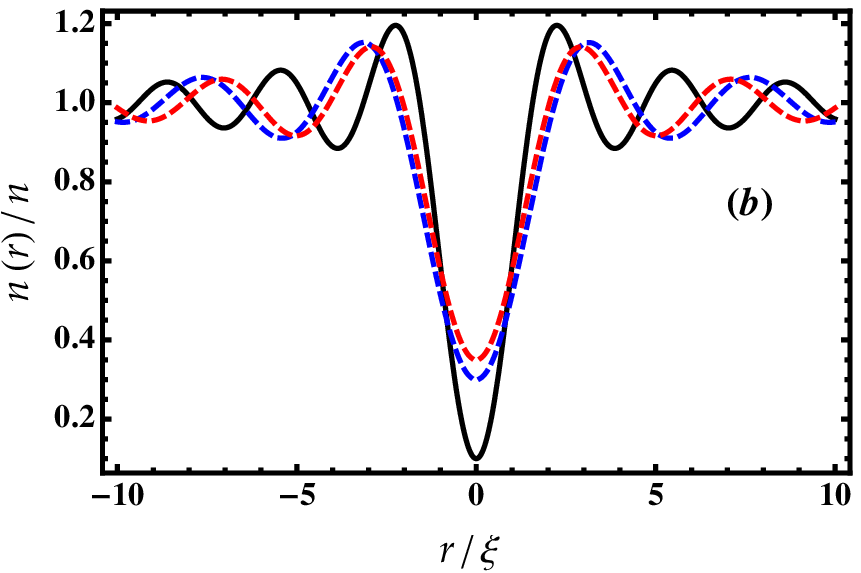}
\caption {Condensed density deformed by a disordered speckle potential for  $\sigma /\xi=3$ (a) and $\sigma/\xi=0.5$ (b).
Parameters are: $ U_0 =0.67 n g$, $a=56 a_0$ and $\epsilon_{dd} =1.17$. Here $a_0$ is the Bohr radius.}
\label{dens}
\end{figure}

To check the validity of the developed perturbative approach, we compare our predictions at first order (\ref{sers1}) and second-order (\ref{sers2})
to the exact numerical solutions of the extended GP equation (\ref{GPE}) for a disordered speckle potential corresponding to the correlation function (\ref{Cordis}).
The results are shown in Fig.\ref{dens}. 
In the case when $\sigma >\xi$, the first-order smoothing excellently agree with the full exact numerical solution of Eq.(\ref{GPE}),
while the second-order solution slightly differs from the simulated results (see Fig.\ref{dens}.a). 
This guarantees the validity of the perturbation theory and confirms the condition (\ref{VC}). 
For $\sigma \simeq 0.5\,\xi$, the numerical solutions significantly diverges from the pertubative calculations (see Fig.\ref{dens}.b).
This is most likely due to the strong density modulations introduced by the disorder potential.
Therefore, the perturbation approach is no longer valid in this regime. 
In both cases, the density is depleted near the center.  
This depleted contribution comes from the special form of the disorder potential (\ref{Cordis}).

Let us now calculate the glassy fraction.
Substituting the function (\ref{CorrS}) into Eq.(\ref{depdis}) and integrating over the momentum from 0 to $1/\sigma$,  we obtain
\begin{equation} \label{depdis1}
{n_R}= n_{\text{HM}} h\left(\gamma, \frac{\sigma}{\xi}\right),
\end{equation}
where 
$n_{\text{HM}}=\left(m^2 R_0/8\pi^{3/2} \hbar^4 \right) \sqrt{n/a}$  is the seminal Huang-Meng result for BEC with short-range interaction \cite {HM}. 
The anisotropic disorder function is given as
\begin{equation} \label{func}
h\left(\gamma, \frac{\sigma}{\xi}\right)=  \sqrt{\frac{\gamma}{\epsilon_{dd}}} \int_0^\pi d\theta\frac{\sin\theta S(\alpha)}{\sqrt{1+\gamma (3\cos^2\theta-1)}}, 
\end{equation}
where the parameter $\gamma= \epsilon_{dd}/[1+3 g_{\text{LHY}} n^{1/2}/2g]$,
and $S(\alpha)=\sqrt{\alpha/4\pi^2} [4-(16\alpha+6)\ln\left(1+1/4\alpha\right)+2\sqrt{1/\alpha}\arctan \left(1/\sqrt{4\alpha}\right)]$
where $\alpha=\sigma^2[1+\gamma (3\cos^2\theta-1)]/\xi^2$.   
The disorder function (\ref {func}) is important since it explains the interplay between the disorder potential, the LHY quantum corrections and the DDI.
For $g_{\text{LHY}}=0$, $\gamma= \epsilon_{dd}$, thus,  the function (\ref {func}) reduces to that obtained for a dipolar BEC in weak isotropic speckle disorder \cite{Boudj}.
The main difference between the function $h(\gamma, \sigma/\xi)$ and that found in our recent work\cite{Boudj} is that the former remains finite even for $\epsilon_{dd} >1$, 
while the later diverges in the limit $\epsilon_{dd} >1$ due to the dipolar instability.
For $\sigma/\xi\rightarrow 0$ and $\epsilon_{dd} =g_{\text{LHY}}=0$, we read off from Eq.(\ref{func}) that one obtains $h\left(\gamma, \frac{\sigma}{\xi}\right) \rightarrow 1$. 
Therefore, we accurately recover the Huang-Meng result \cite{HM}. 
When $\epsilon_{dd} =0$, we reproduce the results of \cite {Abdulaev} for a nondipolar condensate with isotropic laser speckle random potential.

\begin{figure}[htb] 
\includegraphics[scale=0.58]{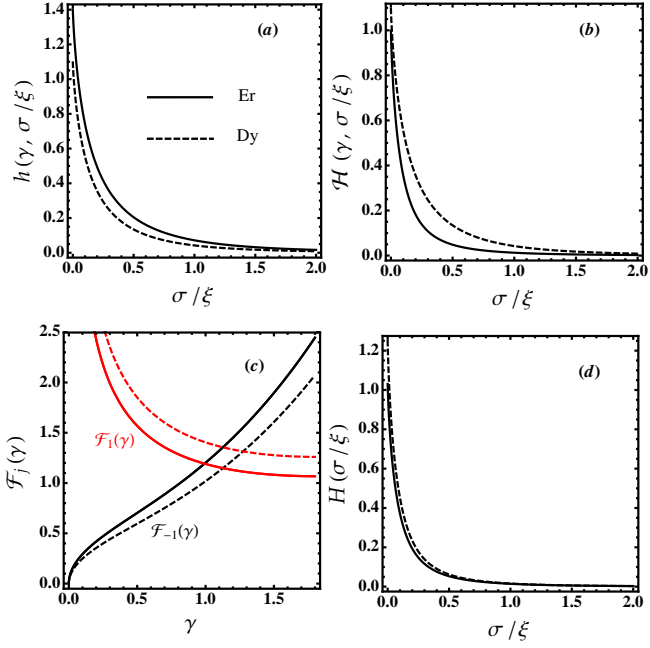}
\caption {Different disorder functions as a function of $\sigma/\xi$ for laser speckle potential. 
Paramters are :  $a=80 a_0$, $\epsilon_{dd} = 1.63$, $n \sim 4 \times 10^{21}$ m$^{-3}$, for ${}^{164}$Dy atoms \cite {Pfau3}, 
and $a=56 a_0$, $\epsilon_{dd} =1.17$, $n \sim 35 \times 10^{20}$ m$^{-3}$ for ${}^{166}$Er atoms \cite {Chom}.}
\label{disF}
\end{figure}


Figure \ref{disF}.a shows the effects of the disorder correlation length on the behavior of the disorder function.
Many important results presented in Fig.\ref{disF}.a should be noted.
First, the glassy fraction ${n_R}/n_{\text{HM}}$ described by the function $h$ decreases with the disorder correlation length.
A surprising result is that the glassy component ${n_R}$ lowers as the dipolar interaction goes stronger notably for $\sigma <\xi $.
This can be interpreted as the fact that the LHY quantum corrections suppress the DDI effects
and hence, the fragments coalesce in a single extended condensate which restores superfluidity.
This is in stark contrast to the disordered dipolar BEC without LHY fluctuations where the DDI tend to strongly increase the disorder fraction \cite{Boudj,Nik,Krum}.

In the limit of a delta-correlated disorder where $\sigma/\xi\rightarrow 0$, the glassy fraction
reduces to  ${n_R}/n_{\text{HM}}= {\cal F}_{-1} (\gamma)$, where ${\cal F}_{-1} (\gamma)=\sqrt{\gamma/\epsilon_{dd}} {\cal Q}_{-1}(\gamma)$. 
In such a situation, the LHY corrections can significantly enhance the glassy fraction inside the condensate (see Fig.\ref{disF}.c).
As in the case of the laser speckle potential, $n_R$ grows with diminishing $\epsilon_{dd}$ (${\cal F}_{-1} (\text{Er})> {\cal F}_{-1} (\text{Dy}))$.





The correction of the condensate EoS due to the disorder fluctuations can be computed using the perturbative calculation presented in Eq.(\ref{EoS}).
Then, after some algebra, we find  
\begin{equation}\label {chem}
\delta \mu=3 g n_{\text{HM}} {\cal H} \left(\gamma, \frac{\sigma}{\xi}\right),
\end{equation}
where the disorder function is defined by
${\cal H} \left(\gamma, \sigma/\xi\right)= \sqrt{ \epsilon_{dd}/\gamma} \int_0^\pi d \theta \sin\theta \sqrt{1+\gamma (3\cos^2\theta-1)} S(\alpha)$,
its behavior is displayed in Fig.\ref{disF}.b. 
We see that for a large disorder correlation length, the contribution of the disorder on the EoS is not important. 
A careful analysis of the same figure reveals that at fixed $\sigma/\xi$, $\delta \mu$ increases with increasing the DDI.
In the case of delta-correlated disorder, $\delta \mu=3 \,g n_{\text{HM}} {\cal F}_1 (\gamma)$, where ${\cal F}_1 (\gamma)=\sqrt{ \epsilon_{dd}/\gamma}\, {\cal Q}_1(\gamma)$. 
It decreases with $\gamma$ until it reaches its minimal value as is seen in Fig.\ref{disF}.c.
For $g_{\text{LHY}}=0$, the EoS (\ref{chem}) coincides with our EoS obtained recently for a dipolar BEC without LHY term \cite {Boudj}.


We now look at how the external random potential modifies the compressibility.
A straightforward calculation employing Eq.(\ref{Comp}) yields a useful expression for the compressibility shift due to the disorder potential
\begin{equation}\label {Comp1}
\frac{\delta \kappa^{-1}}{ n^2} =\frac{3g} {2n} n_{\text{HM}}  H \left(\gamma, \frac{\sigma}{\xi}\right),
\end{equation}
where the function $H \left(\gamma, \sigma/\xi \right)=  {\cal H} \left(\gamma, \sigma/\xi \right) +(3\beta/64)  h \left(\gamma, \sigma/\xi \right)$,  
and $\beta=(3/4) \sqrt{na^3/\pi} {\cal Q}_5 (\epsilon_{dd})$.\\
Figure \ref{disF}.d. shows that the function $H$ of the inverse compressibility lowers with rising the disorder correlation length $\sigma$,
while it increases with increasing $\epsilon_{dd}$.
Notice that the sound velocity is related to the inverse compressibility, where the decrease in $\kappa^{-1}$ leads to decrease the sound velocity and vice versa.

We now proceed with the perturbative calculation of the superfluid density. 
Depending on the boost direction, $n_s$ splites into two different directions parallel or perpendicular to the dipole orientation. \\
In the parallel direction, the superfluid density can be obtained via (\ref{sup1})
\begin{equation}\label {supflui1}
n_s^{\parallel}=n- 4 n_{\text{HM}} h^{\parallel} \left(\gamma, \frac{\sigma}{\xi}\right),
\end{equation}
where the disorder function is given by
$h^{\parallel} \left(\gamma, \frac{\sigma}{\xi}\right)= \sqrt{\gamma/\epsilon_{dd}}\int_0^\pi d\theta [\sin\theta\cos^2\theta S(\alpha)/\sqrt{1+\gamma (3\cos^2\theta-1)}]$. \\
According to Eq.(\ref{sup2}), the superfluid in the perpendicular direction reads
\begin{equation}\label {supflui2}
n_s^{\perp}=n-2n_{\text{HM}} h^{\perp} \left(\gamma, \frac{\sigma}{\xi}\right),
\end{equation}
where $h^{\perp}\left(\gamma, \frac{\sigma}{\xi}\right)=h\left(\gamma, \frac{\sigma}{\xi}\right) -h^{\parallel} \left(\gamma, \frac{\sigma}{\xi}\right)$.

\begin{figure}[htb] 
\includegraphics[scale=0.58]{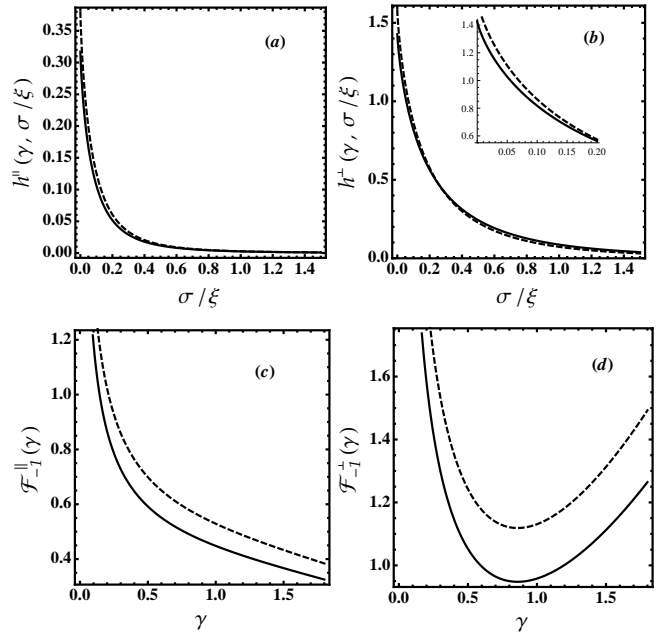}
\caption {Superfluid disorder functions vs. $\sigma/\xi$ for laser speckle potential. 
Parameters are the same as in Fig.\ref{disF}.}
\label{disF1}
\end{figure}

Figure \ref{disF1}.a  depicts that the function $h^{\parallel}$ is slightly increasing with $\epsilon_{dd}$ for fixed $\sigma/\xi$ 
signaling that the DDI play a minor role in the behavior of $n_s^{\parallel}$.
We see also that when the external random potential correlation length becomes larger then the heanling length,
the normal component of the superfluid is diminished ($h^{\parallel}$ is small) which means that the whole liquid becomes practically superfluid in the parallel direction.
This effect can be understood by the fact that in the limit $ \xi < \sigma$, 
the kinetic energy term is small and the  wavefunction of the condensate simply follows the spatial modulations of the potential.
Hence, the dipolar bosons will not localize anymore.

Figure \ref{disF1}.b depicts that in the perpendicular direction, the superfluidity changes its behavior from small to large disorder correlation length $\sigma$.
For $\sigma <0.2 \xi$, $h^{\perp}$ is decreasing with $\gamma$ while for $\sigma >0.2 \xi$, it is increasing function with $\gamma$.
This unusual behavior arises from the competition between the DDI,  the LHY corrections and the disorder potential.
An important remark is that $h^{\parallel} < h^{\perp}$ whatever the value of $\sigma$, indicating that the localized particles can not contribute to superfluidity and hence, form
obstacles for the superfluid flow in the perpendicular direction. 


In the case of a delta-correlated disorder ($\sigma/\xi\rightarrow 0$),  the superfluid density 
in both parallel and perpendicular directions turns out to be given, respectively as  $n_s^{\parallel}=n- 4 n_{\text{HM}} {\cal F}^{\parallel} (\gamma)$,
and $n_s^{\perp}=n-2n_{\text{HM}} {\cal F}^{\perp} (\gamma)$, 
where ${\cal F}^{\parallel} (\gamma)=\sqrt{ \epsilon_{dd}/\gamma}  {\cal Q}^{\parallel}_{-1}(\gamma)$ and 
${\cal F}^{\perp} (\gamma)= \sqrt{ \epsilon_{dd}/\gamma}  {\cal Q}^{\perp}_{-1}(\gamma)$. The function 
${\cal Q}_{-1}^{\parallel} (\gamma)=\frac{1}{3} (1-\gamma)^{1/2} {}_2\!F_1\left(-\frac{1}{2},\frac{3}{2};\frac{5}{2};\frac{3\gamma}{\gamma-1}\right)$, 
features by ${\cal Q}_{-1}^{\parallel} (\gamma=0)=1/3$ and imaginary for $\gamma>1$, and ${\cal Q}^{\perp}_{-1}(\gamma)={\cal Q}_{-1}(\gamma)-{\cal Q}^{\parallel}_{-1}(\gamma)$. 
Figure \ref{disF1}.c shows that the function ${\cal F}^{\parallel}$ is decreasing with $\gamma$ pointing out
that the superfluid fraction is rised in the parallel direction.
Whereas, the function ${\cal F}^{\perp} (\gamma)$  decreases with increasing $\gamma$ until it reaches its minimum at $\gamma =\gamma_c\simeq 8.5$,
while it augments for large values of $\gamma$ as is seen in Fig \ref{disF1}.d. 
So, at a certain critical value $\gamma_c$, $n_s^{\perp}$ becomes significant regardless of the DDI or LHY strengths.


One can infer from Eqs.(\ref{depdis1}), (\ref{supflui1}) and (\ref{supflui2}) that when a Bose condensate is subjected to the action of an external disorder potential,
the relation between condensed and superfluid fractions may be changed owing to the intriguing role of the LHY quantum fluctuations as we have foreseen above. 
For instance, in the case of a BEC with the LHY quantum term, the condensed fraction could be larger or smaller than
the superfluid fraction depending on the flow direction,  whereas for a disordered dipolar BEC without LHY corrections, 
the condensed fraction is always smaller than the superfluid fraction \cite{Boudj, Boudj1}.

\section{Conclusion and outlook} \label{Concl}

In this paper we developed a perturbative theory and analytically studied the effects of the LHY quantum corrections
on the properties of a disordered dilute dipolar BEC at zero temperature. 
Conditions of validity of such a treatment has been well specified.
We found in particular that, in the presence of the LHY, the density of the dipolar condensate 
is basically insensitive to the disorder potential in the regime when $\sigma > \xi$.
Useful analytical expressions for the condensed fracion, the equation of state, the compressibility and the superfluid density have been derived. 
The obtained formulas can be used to experimentally control effects of disorder on a dipolar BEC. 
We compared the relative change of the condensate deformation and of thermodynamic quantities due to disorder with those reported previously in the literature.
We found that the glassy fraction inside the condensate is reduced due to the remarkable role of the LHY quantum fluctuations.
Importantly, the superfluidity changes its properties from correlated to uncorrelated disorder potential.
This unconventional behavior is the result of competition between the LHY corrections, the disorder potential and the DDI.
We showed in addition, that the presence of the DDI renders the perpendicular component of the superfluid density important and highly anisotropic. 
Such an anisotropy which should generate an anisotropic sound velocity can be  experimentally measurable.
A natural generalization of this work is the calculation of the depletion of a disordered Bose gas which can be evaluated by means of the Bogoliubov theory 
\cite{Gaul, Gaul1, Ghab, Boudj, Boudj1,Boudj2, Boudj3}.
A future work toward understanding the effects of an external disorder potential  on the superfluidity and the thermodynamics of a quantum self-bound droplets 
will be also of particular interest to the cold atomic physics community.

\section {Acknowledgments}
We would like to thank Axel  Pelster,  Laurent Sanchez-Palencia and Cord M\"uller for fruitful discussions and insightful comments on the manuscript.

\end{document}